# A proof Procedure for Testing Membership in Regular Expressions

Keehang Kwon and Hong Pyo Ha
Dong-A University Department of Computer Engineering
Busan, Republic of Korea

Jiseung Kim
Kyung-IL University Department of Industrial Engineering
Daegu, Republic of Korea

**Abstract**

We propose an algorithm that tests membership for regular expressions and show that the algorithm is correct. This algorithm is written in the style of a sequent proof system. The advantage of this algorithm over traditional ones is that the complex conversion process from regular expressions to finite automata is not needed. As a consequence, our algorithm is simple and extends easily to various extensions to regular expressions such as timed regular expressions or regular languages with the intersection.

## 1. Introduction

Since its introduction, regular expressions [2,5] have gained much interest for applications such as text search or compiler components. One important question is, given a string *w* and a regular expression *r*, to decide whether *w* is in the set denoted by *r*. Testing membership in a regular expression has traditionally concentrated on converting a regular expression to finite automata. Such a conversion technique is unsatisfactory for at least two reasons. First, the conversion process itself requires a lot of extra overloads. Second, the conversion technique has only a limited number of applications and does not extend well to various – even simple – extensions (regular expressions with time [1], regular expressions with intersection, etc) to regular expressions.

Little has been studied about the algorithms for testing membership for regular expressions themselves. This paper introduces such an algorithm. It is simple, easy to understand, nondeterministic and has some resemblance to the proof theory of intuitionistic linear logic[3] . In addition, it is a simple matter to observe that this algorithm extends well to other extensions to regular expressions.

In this paper we present our algorithm, show some examples of its working, and discuss further improvements. The remainder of this paper is structured as follows. We describe our algorithm in the next section.

In Section 3, we present some examples. Section 4 concludes the paper with some considerations for further improvements.

## 2. The language

The regular expression is described by *r*-formulas given by the syntax rules below:

$$r ::= \emptyset \mid \varepsilon \mid a \mid r \cdot r \mid r + r \mid r^*$$

In the rules above, an alphabet *a* represents the set {*a*}. ε represents {ε}. Ø represents the empty set. *r* · *s* represents the concatenation of two sets *r* and *s*. *r* + *s* represents the union of *r* and *s*. The Kleene closure of *r* - *r\** - indicates there are any number of *r*.

We often write *rr* in place of *r* · *r*. The degree *d(r)* of a regular expression *r* is defined as follows: $d(\emptyset)=0$, $d(a)=1$, $d(r \cdot s) = d(r+s) = d(r) + d(s) + 1$, $d(r^*) = d(r) + 1$. Further, $d(r_1,\ldots,r_n) = d(r_1)+\ldots+d(r_n)$ where $r_1,\ldots,r_n$ is a sequence of regular expressions.

The question of whether a string is a member of a regular expression is quite interesting. Such an algorithm needs to cope with the following: (1) the associativity of the operators, e.g., $abc \in a \cdot (b \cdot c)$, $abc \in (a \cdot b) \cdot c$, and (2) the multiplicity of the Kleene closure, *e.g.*, $aaa \in a^* \cdot a^*$, in an elegant fashion. We will present an algorithm for this task in the style of a proof system.

Let $w$ be a regular expression and $r_1,\ldots,r_n$ be a list of regular expression. Then a *sequent* of the form $r_1,\ldots,r_n \vdash w$ – the notion that is an element of the concatenations of $r_1,\ldots,r_n$ – is defined constructively by two axioms and eight inference rules. This is shown below.

Algorithm for Testing Membership

$$\frac{}{a \vdash a} Axiom1 \qquad \frac{}{\vdash \epsilon} Axiom2$$

$$\frac{\rho, \Psi, \Delta \vdash w}{\rho \cdot \Psi, \Delta \vdash w} \cdot L \qquad \frac{\Delta_1 \vdash w_1 \quad \Delta_2 \vdash w_2}{\Delta_1, \Delta_2 \vdash w_1 w_2} \cdot R$$

$$\frac{\Delta \vdash w}{\rho, \Psi \vdash w} \epsilon L \qquad \frac{\Delta \vdash w}{\rho^*, \Delta \vdash w} WL$$

$$\frac{\rho^*, \rho^*, \Delta \vdash w}{\epsilon, \Delta \vdash w} CL \qquad \frac{\rho, \Delta \vdash w}{\rho^*, \Delta \vdash w} DL$$

$$\frac{\rho, \Delta \vdash w}{\rho + \Psi, \Delta \vdash w} + L_1 \qquad \frac{\Psi, \Delta \vdash w}{\rho + \Psi, \Delta \vdash w} + L_2$$

In the above rule, $a$ is an alphabet, $\Delta$ denote a list of regular expressions and $\rho, \psi$ denote a single regular expression. An inference rule can be read as follows: if all the sequents above are true, then the sequent below is true. A sequent $\Delta \vdash w$ has a *proof* if $\Delta \vdash w$ can be obtained from the axioms by applying the inference rules. In dealing with $\rho^*$ construct, the proof system can either discard it, use $\rho$ once, or use $\rho$ at least twice.

Let us refer to the above collection of axioms and inference rules as *DS*. The following theorem shows the sound and completeness of the proof system *DS*.

**Theorem 2.1:** Let $r_1,\ldots,r_n$ be a list of regular expressions and let $w$ be a string. Then, $w$ is an element of $r_1\ldots r_n$ if and only if $r_1,\ldots, r_n \vdash w$ has a proof in *DS*.

*Proof.* The reverse direction is straightforward. In the forward direction, we prove the theorem by an induction on the degree of the sequence $r_1\ldots, r_n, w$. If the degree is 1 or 2, then it must be of the form $\varepsilon \in \varepsilon$, $\varepsilon \in$, or $a \in a$ where $a$ is an alphabet. It is easy to see that the theorem is true.

If the degree is greater than 2, we consider the cases for the structure of $r_1$.

If $r_1$ is $a$, then it must be the case that $w$ is of the form $aw'$ and $w' \in r_2,\ldots, r_n$ where $w'$ is a (possibly empty) string. By the hypothesis, $r_2\ldots, r_n \vdash w'$ has a proof. Putting the proofs for $a \vdash a$ and $r_2\ldots r_n \vdash w'$ using a $\cdot R$ rule, we obtain a proof satisfying the theorem.

If $r_1$ is $r+s$, then it must be the case that $w \in r, r_2\ldots r_n$ or $w \in s, r_2\ldots, r_n$. Consider the former case. By the hypothesis, $r, r_2,\ldots r_n \vdash w$ has a proof. Putting this proof using a $+L_1$ rule, we obtain a proof satisfying the theorem. The same argument can be supplied for the latter case.

The arguments for $rs$ follow a similar pattern. For the case that $r_1$ is of the form $r^*$, we have to consider the three cases depending on whether $r$ is never used, used once, or used more than once. Consider the case where $r$ is never used. Then $w$ must be an element of $r_2,\ldots, r_n$. By the hypothesis, $r_2,\ldots, r_n \vdash w$ has a proof. Putting this proof using a *WL* rule, we obtain a proof satisfying the theorem. This is shown below.

$$\frac{r_2,\ldots, r_n \to w}{r^*, r_2,\ldots, r_n \to w} WL$$

If $r$ is used once, then there must be a string $z$ such that $w = zw'$, $z \in r$, and $w' \in (r_2,\ldots, r_n)$. By the hypothesis, both $r \vdash z$ and $r_2,\ldots, r_n \vdash w'$ have proofs. Putting the proofs for $r \vdash z$ and $r_2,\ldots, r_n \vdash w'$ using the rules, we obtain a proof satisfying the theorem.

This is shown below.

$$\frac{\dfrac{r \to z \quad r_2,\ldots, r_n \to w'}{r, r_2,\ldots, r_n \to zw'} \cdot R}{r^*, r_1,\ldots, r_n \to zw'} DL$$

The arguments for the remaining case follow a similar pattern. The additional observation is that the *CL* rule is needed for this case.

## 3. Examples

This section describes the use of our algorithm. An example is provided by the following proof of $aa \in a^*$.

It is interesting to note that the *CL* rule is used to control the multiplicity of $a^*$.

$$\frac{\dfrac{\dfrac{a \to a}{a^* \to a} DL \quad \dfrac{a \to a}{a^* \to a} DL}{a^*, a^* \to aa} \cdot R}{a^* \to aa} CL$$

Another example of our algorithm is provided by the following proof of the sequent $ca \in (b+c) \cdot a$

$$\cfrac{\cfrac{\cfrac{c \to c}{b+c \to c} + L_2 \quad a \to a}{b+c, a \to ca} \cdot R}{(b+c) \cdot a \to ca} \cdot L$$

A computation process typically searches for a proof from the bottom-up in a sequent calculus for reasons of efficiency. Thus, given a conclusion sequent, it attempts to find its proof from bottom-up.

## 4. Conclusion

We have described an algorithm for testing membership in regular expressions. The advantage of this algorithm is that it does not require the complex conversion process to finite automata. As a consequence, it extends easily to various extensions to regular expressions. For example, our algorithm extends easily to the one that deals with algebraic laws, *i.e.*, regular expressions with variables [5]. Two regular expressions with variables are equivalent if whatever expressions we substitute for the variables, the results are equivalent. For example, $\forall L \forall M(\qquad L+M=M+L \qquad)$.
    Regarding the performance of our algorithm, non-determinism is present in several places of this algorithm. In particular, there is a choice concerning which way the text is split in the ·R *rule*. Hodas and Miller [4] dealt with this rule by using IO-model in which each goal is associated with its input resource and output resource. The idea used here is to delay this choice of splitting as much as possible. This observation leads to a more viable implementation.   Our ultimate interest is in a procedure for carrying out computations of the kind described above. It is hoped that these techniques may lead to better algorithms.

## 5. Acknowledgements

This paper was supported by Dong-A University Research Fund.